# A Transferable Machine-learning Scheme from Pure Metals to Alloys in Predicting Adsorption Energies

*Xin Li, Bo Li, Zhiwen Chen, Wang Gao,\* and Qing Jiang*

Dr. X. Li, Dr. B. Li, Prof. W. Gao, Prof. Q. Jiang
Key Laboratory of Automobile Materials, Ministry of Education, Department of Materials Science and Engineering, Jilin University 130022, Changchun, China
E-mail: wgao@jlu.edu.cn

Dr. Z. Chen
Department of Materials Science and Engineering, University of Toronto, Toronto, ONM5S 3E4, Canada

Abstract: Alloys present the great potential in catalysis because of their adjustable compositions, structures and element distributions, which unfortunately also limit the fast screening of the potential alloy catalysts. Machine learning methods are able to tackle the multi-variable issues but still cannot yet predict the complex alloy catalysts from the properties of pure metals due to the lack of universal descriptors. Herein we propose a transferable machine-learning model based on the intrinsic properties of substrates and adsorbates, which can predict the adsorption energies of single-atom alloys (SAAs), AB intermetallics (ABs) and high-entropy alloys (HEAs), simply by training the properties of transition metals (TMs). Furthermore, this model builds the structure-activity relationship of the adsorption energies on alloys from the perspective of machine learning, which reveals the role of the surface atoms' valence, electronegativity and coordination and the adsorbates' valence in determining the adsorption energies. This transferable scheme advances the understanding of the adsorption mechanism on alloys and the rapid design of alloy catalysts.

## 1. Introduction

Alloys have shown great promise in the field of heterogeneous catalysis because they offer a vast chemical space through different elements and mixing ratios.[1] There are a wide variety of alloys scoping from single-atom alloys (SAAs),[2] bi- or multi-metallic alloys (like AB intermetallics (ABs))[3] to high-entropy alloys (HEAs),[4] each with their own unique catalytic properties. Depending on the chemical compositions, element distributions, and morphologies (like nanoparticles (NPs)), alloy catalysts can present the increased activity and selectivity in comparison to monometallic catalysts due to their multiple functional adsorption sites.[5] Unfortunately, the growing chemical spaces and complicated structures prohibit the screening of the potential alloy catalysts and the understanding of the intrinsic determinants of alloys in adsorption. Therefore, it is essential to build a model to quantify the alloys' intrinsic properties on their adsorption properties.

Many attempts have been made to address this issue like the d-band model,[6] linear scaling relationships[7] and the intrinsic-parameter model,[8] which have been applied to transition metals (TMs) and partial near-surface alloys (NSAs) with some success, but can hardly be extended to the ABs and HEAs. Fortunately, in the growing complexity of structure-property relationships, the data-driven methods such as machine learning (ML) have been employed increasingly to tackle this dilemma.[9] However, the ML methods demand the huge amount of training sets that are commonly obtained from time-consuming DFT calculations. Especially for HEAs with the random element distribution, it is difficult to obtain the large representative datasets. Moreover, most of the recent ML studies aim at one kind of alloys, that is, training and testing on the same kind of alloys. These models need to enlarge the existing datasets for the unknown types of alloys, which lack the universality and transferability between the different kinds of alloys. In consequence, a simple and universal ML framework is urgently needed for quickly estimating the adsorption energies of alloys from the properties of pure metals and for deeply understanding the adsorption mechanism on the different alloy systems.

The key of the ML framework is the descriptors that express the correlation between the adsorption energies and the intrinsic properties of substrates and adsorbates.[10] A typical descriptor is from the d-band model by Nørskov et al, which provides the d-band center to understand the trend of adsorption energies on TM surfaces.[6] However, the d-band center is derived from time-consuming DFT calculations and often cannot satisfy the desired accuracy for alloys.[9f, 11] Generalized coordination number works well for pure metals and nanoparticles,[12] and the modified coordination number can describe TM oxides reasonably.[13] However, the application of these coordination descriptors is limited on the complex alloy surfaces. Clearly, one needs to identify new useful descriptors to capture the intrinsic characteristics of alloys. Coupled with the ML methods, the chosen descriptors should not only provide universal predictions on the adsorption energies from pure metals to complex alloys, but also enable important implications for uncovering the alloying effect and understanding the adsorption mechanism of complex alloys.

Here we propose a ML model based on the intrinsic descriptors of substrates and adsorbates, which poses a good transferability in predicting the adsorption energies of NPs, SAAs, ABs and HEAs only by training the properties of TMs. Our model can even provide an accurate prediction for the perturbation of the adsorption energies caused by the exchange of active-center atoms on HEAs. Furthermore, the model unravels the correlation between the adsorption energies and the electronic and geometric properties of the substrates and adsorbates. Our model serves as an effective tool for the fast estimation of adsorption energies on complicated alloys, providing a solution for rapidly screening alloy catalysts and deeply understanding the adsorption mechanism.

## 2. Results and Discussion

### 2.1. The construction of models

We prepare a database of the adsorption energies containing 506 datasets on TMs, 178 datasets on TM clusters, 281 datasets on SAAs, 258 datasets on ABs, and 1178 datasets on HEAs, with the various surfaces, adsorption sites and adsorbates (More details are illustrated in Figure 1 and Note S1 in the Supporting Information). We take an adsorption site (which is represented by the green empty circles in Figure 1b-d) and its nearest neighboring atoms (the yellow empty triangles in Figure 1b) as the active center of the adsorption. The nearest neighboring atoms at the active center are further classified according to their bonding number to the adsorption-site atoms. Taking the bridge site on fcc(111) as an example (Figure 1c), there are 2 adsorption-site atoms and 13 nearest neighbors at the active center, the latter of which have three atoms bonding with both of the two adsorption-site atoms (the yellow empty triangles in Figure 1c, represented by $Cg_1$) and the remaining ten atoms that bonding with only 1 adsorption-site atom (represented by $Cg_2$). The division of atoms on hollow sites is illustrated in Figure S1.

In our previous study, we have proposed an expression of the adsorption energies ($E_{ad}$) for TMs, NPs, partial NSAs, and TM oxides as follows:

$$E_{ad} = 0.1 \times \alpha \times \psi + 0.2 \times (1 - \alpha) \times \overline{CN} + \theta \quad (1)$$

where $\psi$ and $\overline{CN}$ are the electronic descriptor and generalized coordination number of substrates. $\psi$ is defined as:

$$\psi = \frac{(\prod_{i=1}^{N} S_{v_i})^{\frac{2}{N}}}{(\prod_{i=1}^{N} \chi_i^{\beta})^{\frac{1}{N}}} \quad (2)$$

where $N$ is the number of the atoms at the active center. $S_{v_i}$ and $\chi_i$ are the outer-electron number and electronegativity of the $i$th atom at the active centers. $\beta$ is a parameter determined by the contribution of d-orbitals and sp-orbitals to the valence and electronegativity. $\beta$ is 1/2 for Ag and Au and 1 for the other TMs, reflecting the fact that the d-state contribution to the adsorption is much less important in Ag and Au than in the other TMs.[8] $\overline{CN}$ is obtained by dividing the sum of the nearest-neighbor own usual coordination number $CN$ with the usual $CN$ in the bulk.[12] We compile a Python script to calculate $\psi$ and $\overline{CN}$ in order to obtain a large number of descriptors of the targeted structures conveniently (see more details in Note S2 in the Supporting Information). $\alpha$ is a characteristic parameter for adsorbates. It is defined as $\frac{X_m - X}{X_m + 1}$, where $X_m$ and $X$ are the maximum bondable number and actual bonding number of the central atom for a given adsorbate.[8] In our ML model, we adopt $\psi_1$ and $\psi_0$ for characterizing the electronic structures of the active center and adsorption site, $\overline{CN}$ for representing the geometric structure of substrates and $\alpha$ for doing the properties of adsorbates. Moreover, we introduce the geometric means of $(\psi CN)_{as}$ and $(\psi CN)_{Cg_1}$ for the adsorption-site atoms and the atoms in the category Cg$_1$ respectively, which defined as:

$$(\psi CN)_{as/Cg_1} = ( \prod_{i=1}^{N} (\psi_{s_i} CN_i) )^{\frac{1}{N}} \tag{3}$$

where $N$ is the number of atoms at the adsorption site or the category Cg$_1$. $\psi_{s_i}$ and $CN_i$ represent the electronic descriptors and ordinary coordination number of the $i$th atoms at the adsorption site or the category Cg$_1$. Additionally, the period number $N_p$ of elements is used to account for the difference of elements in the same group.

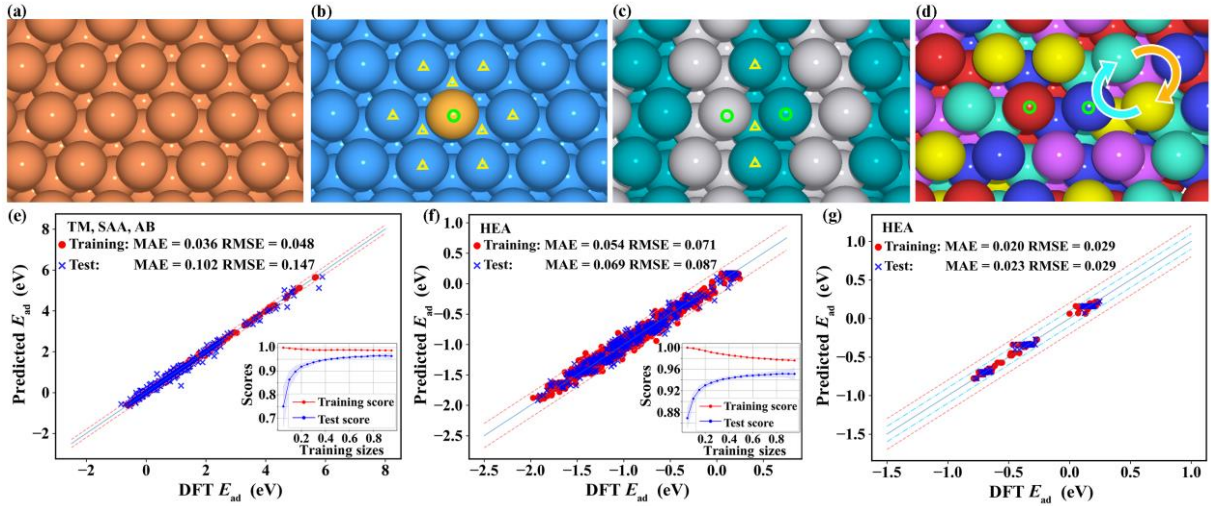

**Figure 1.** Structures of TM (a), SAA (b), AB (c) and HEA (d). Herein, adsorption-site atoms are represented by the green empty circles and the atoms with more bonding numbers with adsorption-site atoms (in category $Cg_1$) are represented by the yellow empty triangles in (b) and (c). (d) also shows the case of the interchange of nearest neighboring atoms on HEAs. Results of cross-validation models for TMs, SAAs and ABs (e) and HEAs (f). The subgraphs show the curves of the scores of the cross-validation model with the training sizes. (g) shows the result of a perturbation on the adsorption energy for the interchange of atoms around the adsorption site. The blue solid line in (e), (f) and (g) indicates that the predicted value is the same as the actual value. The red dashed line in (e), (f) and (g) and the light blue dot dashed line in (g) indicate that the error between the predicted value and the actual value is ±0.2 eV and ±0.1 eV.

To understand the relationship between the intrinsic descriptors and the targeted adsorption energies, we evaluate the proposed descriptors via the Extreme Gradient Boosting (XGBoost) regression algorithm.[14] XGBoost is an extension of Gradient Boosting Decision Tree (GBDT). It shows an overwhelming performance in many fields due to its high efficiency and flexibility. Furthermore, XGBoost is quantified for both classification and regression analysis, which is applicable to our data that covers six kinds of adsorbates and nine kinds of metals as well as their alloys with various adsorption sites. With our proposed descriptors, the XGBoost algorithm not only shows the higher accuracy and better adaptability than the simple algorithms like linear regression in predicting the adsorption energy, but also exhibits the better interpretability than neural networks.

**2.2. Training and testing on the same kind of metallics.**

SAAs are formed by doping the TM surfaces with a single atom, which causes a relatively localized change of the TM surfaces, while ABs are highly periodic and exhibit the relatively non-local alloying effects. From TMs to SAAs and ABs, the surface structures become complex and the values of $\psi$ turn to be relatively continuous. Despite the pronounced difference between TMs, SAAs and ABs, the ML model with the nine proposed descriptors $\psi_0$, $\alpha$, $\alpha\psi_0$, $\overline{CN}$, $CN$, $(1-\alpha)\overline{CN}$, $N_p$, $(\psi CN)_{as}$ and $(\psi CN)_{Cg_1}$ provides a root mean square error (RMSE) of 0.147 eV on the test sets (Figure 1e). Our model performs much better than the ML models based on the d-band center (with the RMSE of 0.37 eV) and the linear scaling relations (with the RMSE of 0.28 eV). The accuracy of our model is similar to Andersen's SISSO prediction model (with the RMSE of 0.15 eV), in which the primary features contain 18 parameters such as the sp- and d-band properties and surface work function that need expensive DFT calculations.[9f, 11] Our proposed descriptors are the intrinsic properties of metallics and can be easily accessible by the table looking up, which are convenient for the practical application.

HEAs, defined as near-equimolar alloys of four or more elements, exhibit multiple compositions and the random ordering of elements, which can generate a large number of potential active sites on the surfaces.[4, 15] However, the complex structures of HEAs also prohibit understanding the underlying alloying effects. With the ML model, we adopt $\psi_0$, $\overline{CN}$, $CN$, $(\psi CN)_{as}$ and $(\psi CN)_{Cg_1}$ to predict the adsorption energies of HEAs. The RMSE is 0.095 eV on the test data, which indicates that the chosen descriptors are sufficient to characterize the adsorption properties of HEAs. Considering that the studied species is adsorbed on the bridge site, we refine the descriptor of $\psi CN$ into the feature that represents each adsorption-site atom and each neighboring atom in category Cg$_1$. The RMSE on the test sets decreases to 0.087 eV (Figure 1f), which is smaller than the RMSE of the neural network model (0.116 eV) with the complex hyperparameters and the descriptors such as the elemental information.[16] Instead of the information of all the atoms at the active centers, one just needs the information of the adsorption-site atoms and the nearest neighboring atoms with more bonding number to the adsorption-site atoms (the neighboring atoms in the category Cg$_1$) in order to predict the adsorption energies of HEAs according to our model. These findings significantly improve the efficiency and interpretability of the ML model.

Note that the random arrangement of atoms in HEAs can result in a diverse behavior of the adsorption energy. A special case for HEAs is that for an adsorption site with the specific atoms, the interchange of its nearest neighboring atoms also leads to a perturbation on the adsorption energy, which is non-negligible but still unpredictable. We calculate the adsorption energies for this case, finding that these permutations lead to an adsorption-energy variation up to 0.3 eV. The proposed descriptors $\psi CN$ for each adsorption-site atom and each atom in category $C_{g1}$ with the ML model are found to distinguish these adsorption energies effectively with a RMSE of 0.029 eV (Figure 1g). In practice, the adsorption energy of *OH on the optimal O$_2$ reduction reaction (ORR) catalysts should be 0.1 eV weaker than that on Pt(111). Our model is thus sufficient to capture the delicate change of the adsorption energies on HEAs to optimize the catalytic activity for ORR.

It is known from the learning curves (the subgraphs of Figure 1e and 1f) that with 20%-data training, the cross-validation score of the test sets is larger than 0.9. Obviously, a small part of data is able to predict a large adsorption-energy dataset efficiently, so that our ML scheme is robust for training and testing on the same metallic systems.

**2.3. The transferability of ML models**

In a regular ML model that trains and tests on the same kind of metallics, the test sets are similar to the training sets and the adsorption characteristics of each metallic system can be learned. For a transferable ML model, the training sets and the test sets are relatively independent. The chosen descriptors need to be universal to capture the similarity between the different systems.

We begin by verifying the transferability of our model in predicting the adsorption energies of SAAs by training the data of TMs (Figure 2a). With the nine intrinsic descriptors $\psi_0$, $\alpha$, $\alpha\psi_0$, $\overline{CN}$, $CN$, $(1-\alpha)\overline{CN}$, $N_P$, $(\psi CN)_{as}$ and $(\psi CN)_{C_{g1}}$, the RMSE on the test sets (SAAs) is 0.157 eV, which is close to the accuracy of the regular ML model (0.147 eV in Figure 1). Among the chosen descriptors, $(\psi CN)_{as}$ and $(\psi CN)_{Cg1}$ represent the contribution of the adsorption-site atoms and the neighboring atoms to adsorption respectively, which are negligible in our transferable model on SAAs. Simply using $\psi_0$, $\alpha$, $\alpha\psi_0$, $\overline{CN}$, $CN$, $(1-\alpha)\overline{CN}$ and $N_p$ without including $(\psi CN)_{as}$ and $(\psi CN)_{Cg1}$, the adsorption energies of SAAs can be estimated from the data of TMs with the RMSE of 0.156 eV. The feature importance (the subgraph of Figure 2a) also suggests that $\alpha\psi_0$ contributes much more to the adsorption energy than all the other descriptors. These results demonstrate that the electronic effect of the active centers on SAAs is highly localized, reflecting a small perturbation of the host on the electronic structures of the dopant. Our model also supports the findings by Greiner et al[17] that demonstrate the free-atom-like electronic structures on the dopants of SAAs. This consistency further confirms the reliability and generality of our transferable model from TMs to SAAs. One needs to mainly focus on the elements of the adsorption sites (namely the values of $\alpha\psi_0$) in the selection of the optimal SAA catalysts with the ML methods.

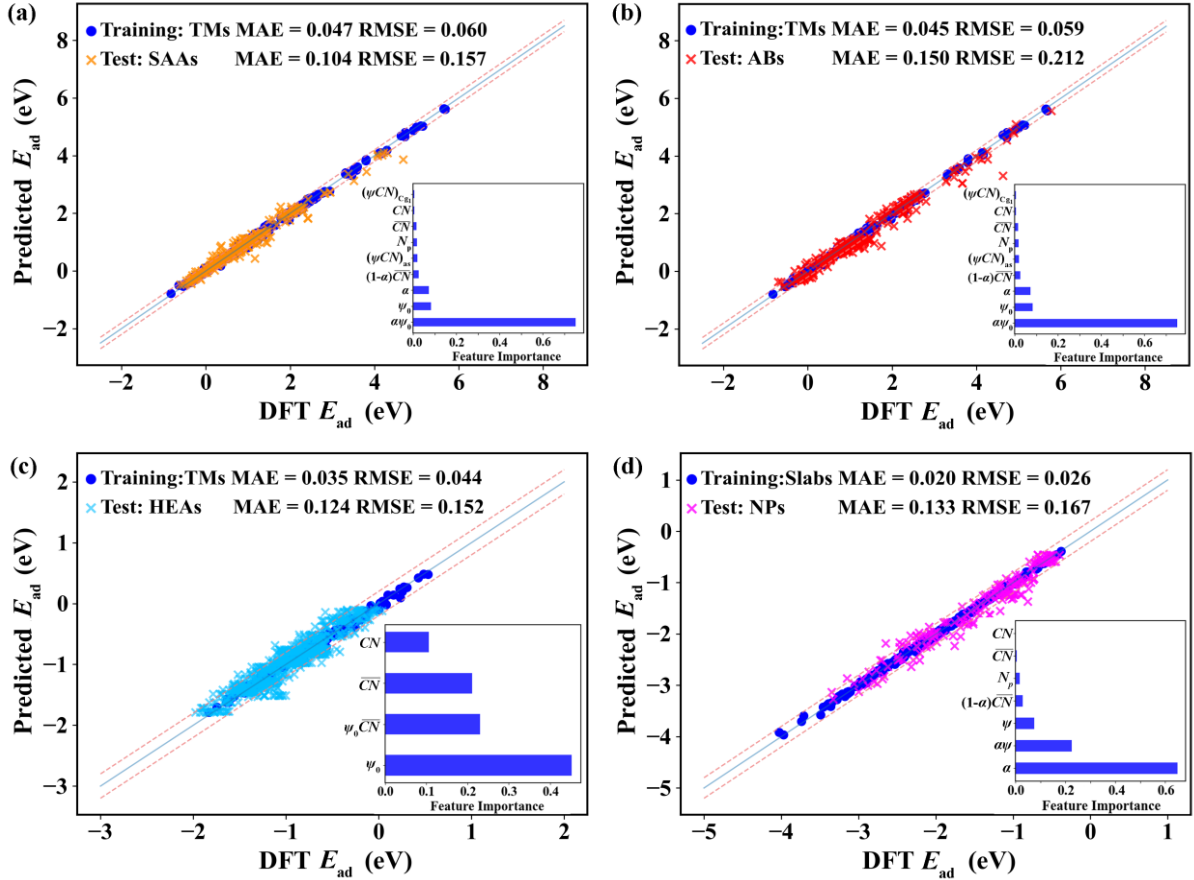

**Figure 2**. The transferability of our models in predicting the adsorption energies of SAAs (a), ABs (b) and HEAs (c) with various adsorption sites, facets and adsorbates by training the data of TMs, as well as the adsorption energies of NPs by training the data of TM surfaces (d). The subgraphs show the importance for the chosen descriptors.

Figure 2(b) shows the transferability of our model between TMs and ABs. Only the data of seven TMs, Ag, Au, Pd, Pt, Ir, Rh and Ru, are learnt to predict the adsorption energies of AgPd, AgAu, PtRh and IrRu by using the nine intrinsic descriptors. The mean absolute error (MAE) of the test sets (ABs) is 0.150 eV, indicating that most of the predicted data are within the accuracy of DFT calculations. This accuracy is on par with the state-of-the-art adsorption prediction models, including the convolutional graph neural network model (MAE: 0.15 eV[18]) and the smooth overlap of atomic position model (SOAP, MAE: ~0.12 eV[19]) that are trained by a large number of the data on bimetallic alloys. ABs are highly periodic and exhibit relatively non-local effects in adsorption compared with SAAs, which suggests that only the characteristics of the adsorption sites are insufficient to correlate the properties of TMs with ABs and the coupling term $(\psi CN)_{as}$ and $(\psi CN)_{Cg_1}$ cannot be excluded. Considering that the RMSE of 0.212 eV is about the accuracy of DFT calculations, our transferable ML model has largely captured the general characteristics from TMs to ABs based on the nine intrinsic descriptors, and further utilizing other independent descriptors with the non-local electronic effect may improve this accuracy.

To further demonstrate the transferability of our model, we turn to study the more complicated alloys, HEAs. Figure 2(c) shows that the RMSE for predicting the adsorption energies of HEAs is 0.152 eV by adopting $\psi_0$, $\overline{CN}$, $CN$ and $\psi_0\overline{CN}$, even we train the data of TMs from Ref [11] and test the data of HEAs from Ref [16]. It's noteworthy that $\alpha$ and its coupling terms are not used in the model of TMs and HEAs because there is only one adsorbate

OH in the available data of HEAs. Surprisingly, our model characterizes the electronic contribution of HEAs to the adsorption energies by only using the adsorption-site electronic descriptor $\psi_0$ and its coupling terms, suggesting that the electronic effect of the active center on HEAs is highly localized as that on SAAs. These findings are consistent with the study by Lu et al,[16] which finds the contributions of the adsorption-site atoms to the adsorption energies are one order of magnitude higher than those of the nearest neighboring atoms.

Besides the transferability between TMs and alloys, our model also exhibits the transferability between the different sizes of NPs.[20] We find that the ML model based on the descriptors $\psi$, $\alpha$, $\alpha\psi$, $\overline{CN}$, $CN$, $(1-\alpha)\overline{CN}$ and $N_p$ is able to predict the adsorption energies of 172-atoms NPs by training those of 55- and 147-atoms with the RMSE of 0.148 eV (Table 1 and Figure S2a). If the model is only trained on the 45 datasets of 55-atoms NPs, the RMSE on the test data with 147- and 172-atoms NPs is 0.218 eV. Our model also exhibits the transferability between NPs and TM surfaces. The model is trained on the data of the TM surfaces calculated by RPBE functional[21] and tested against the data of the NPs calculated by PBE functional. The RMSE is 0.167 eV on test sets (Figure 2d), which confirms the transferability of our model in predicting the adsorption energies even with the data obtained from the different functionals. These findings help to solve the dilemma in calculating the large-size NPs for DFT methods. The order of feature importance on the TMs with adatoms is different from those on TMs, SAAs, and ABs (the subgraphs of Figure 2) probably because there are three adsorbates $CH_3$, CO and OH on TMs with adatoms but six adsorbates on TMs, SAAs, and ABs. Nevertheless, this difference hardly affects the transferability of our model since for each adsorbate our model captures the correlation between the adsorption energies and the electronic and geometric structures of substrates. Moreover, our model is also able to do the mutual predictions between the NPs with the three morphologies (Cube, Cuboctahedron and Icosahedron), based on the descriptors $\psi$, $\alpha$, $\alpha\psi$, $\overline{CN}$, $CN$, $(1-\alpha)\overline{CN}$ and $N_p$ (see Table 1 and Figure S2). When the model is trained on the data of one morphology, it still captures the properties of the other morphologies with the good accuracy. The RMSE is 0.194 eV, 0.172 eV and 0.151 eV for the "leave-one-in" tests (see Table 1 and Figure S2), in comparison to the accuracy of the model by Deans et al.[20] Moreover, the involved training datasets only contain 50~100 data (see Table 1 and Figure S2d-f), indicating that our model is promising in doing the small-sample-datasets based prediction. These results show that our model can capture the size- and morphology-dependent effect of the adsorption energies on NPs effectively, even with the small datasets.

Overall, we have verified the transferability of our ML scheme in predicting the adsorption energies between the different metallic systems by using the intrinsic and accessible descriptors. Our model allows the fast estimation of the adsorption energies for SAAs, ABs and HEAs only based on the properties of TMs and captures the size- and shape-dependent effects of NPs on the adsorption energies, which can greatly accelerate the search and design of the potential alloy catalysts.

**Table 1.** The accuracy of the "leave-one-in" tests on the NPs with different sizes and morphologies along with that by Dean et al.

| Training set | Training size | The RMSE of the test set (eV) | The RMSE of the test set by Dean et al (eV)[20] |
|---|---|---|---|
| 55,147-atoms | 123 | 0.148 | - |
| 55-atoms | 45 | 0.218 | 0.192 |
| Cube | 54 | 0.194 | 0.203 |
| Cubectahedron | 72 | 0.172 | 0.186 |
| Icosahedron | 51 | 0.151 | 0.181 |

**2.4. The analysis and evaluation of models.**

We further understand the similarities and differences between SAAs, ABs and HEAs from our transferable model. $\psi_0$ and its coupling terms, which represent the contribution of the adsorption sites, always play the most important roles in predicting the adsorption energies. This reflects that the electronic effect of the active sites on adsorption is highly localized on these alloys even down to the adsorption sites. It's noteworthy that the electronic effect on SAAs and HEAs is more localized than that on ABs, since the contribution of the nearest neighboring atoms is almost negligible in predicting the adsorption energies on SAAs and HEAs. Compared with the electronic effect, the geometric effect on the three kinds of alloys exhibits a relatively non-local effect since $\overline{CN}$ includes the effects of the adsorption-sites atoms and the nearest neighboring atoms. For the substrates, their electronic effect contributes more to the adsorption energies than the geometric effect, which corresponds to equation (1). Although the coefficients of the electronic and geometric terms are similar in our adsorption-energy expression (with the value of 0.1 and 0.2), the values of $\psi$ (30~90) are much larger than those of $\overline{CN}$ (3~9). The similar electronic and geometric structures between SAAs, ABs and HEAs determine the transferability of our ML model.

Our proposed descriptors exhibit clear physical meanings, which lay the foundation for the transferability of our ML model. The adsorbates' descriptor $\alpha$ reflects the valence of adsorbates and is universal for small molecules. The values of our electronic descriptor $\psi$ are discrete for pure metals while relatively continuous for alloys. $\psi$ exhibits a rough linear relation with the adsorption energy on both TMs and alloys. Therefore, the XGBoost regressor is able to establish the model from the discrete data that are applicable to the continuous data based on our descriptors, so that our model can be applied to the data of alloys accurately by training the data of TMs. In comparable, $\overline{CN}$ also exhibits a rough linear relation with the adsorption energy on the NPs with different sizes and morphologies, so that the model exhibits transferability between the NPs.

In our scheme, the electronic descriptor is obtained with the geometric mean of the adsorption-site atoms for alloys, which expresses the electronic overlap of two or more different elements at the adsorption sites. In the previous studies, the effective mapping between the adsorption energies and the surface properties is established by using the information of every atom at the active center as descriptors.[9f, 16] These models are usually confined to the data that include only one kind of adsorption site to ensure the constant number of descriptors. Taking the bridge sites as an example, each adsorption-site atom is represented by an individual descriptor. These ML models capture the feature of only one adsorption-site atom because the values of adsorption-site atoms are all the same on TMs, resulting in a loss of the properties of the remaining adsorption-site atom on alloys. Moreover, these models cannot be extended to the data with multiple adsorption sites because the number of descriptors cannot be unified between the different adsorption sites. The descriptor $\psi$ includes the contribution of all the adsorption-site atoms to avoid this problem. Therefore, our model exhibits a good performance of transferability in predicting the adsorption energies of SAAs, ABs and HEAs from the properties of TMs.

To verify the validity of our portable ML model, we predict the adsorption energies of the alloys beyond the available datasets. Since Cu is unique among the TMs in catalyzing $CO_2$ reduction reaction ($CO_2$RR),[9c, 22] we select the alloys of which the descriptors are similar to that of Cu. Most of the Cu-doped SAAs on facet(211) satisfy this criterion and their adsorption energies differ little (the maximum difference is 0.12 eV), which confirms the localized electronic effect of the active centers on SAAs. Here we recommend Cu@Ag and Cu@Au, because the adsorption of CO is stronger on Cu than that on the host, so that the adsorbate is easy to be captured by the Cu-doped atom. For ABs and HEAs, CuAg, CuAu, CuIr, CuPt, CuPd, CuRh and AgAuCuPtPd with Cu as the top site meet the requirements. Note that all the predictions are only based on the training data of TMs, and the predicted adsorption energies of these candidates are illustrated in Table S1. Figure 3 shows the free energy diagram of $CO_2$RR on Cu@Ag, CuAg and AgAuCuPtPd surface. The three alloy catalysts produce CO with a limiting potential of -0.47 V, -0.39 V and -0.59 V. Compared with Au for CO production (with the onset potential of -0.50 V by the DFT calculations and -0.68 V by the experiments),[23] these alloys exhibit potential advantages in $CO_2RR$.

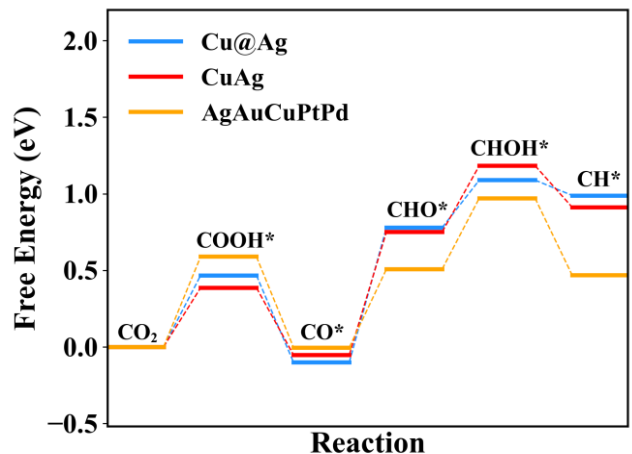

**Figure 3.** The free energy diagram on (211) of Cu@Ag and CuAg as well as (100) of AgAuCuPtPd.

## 3. Conclusion

In summary, we have built a transferable ML model based on the intrinsic descriptors of substrates and adsorbates for rapidly predicting the adsorption energies on metallics. Our model not only enables one to estimate the adsorption energies of SAAs, ABs and HEAs only by training the data of TMs, but also captures the size- and morphology-effect of the adsorption energies on NPs effectively. Our model also captures the delicate variation of the adsorption-energy perturbation with the exchange of active-center atoms on HEAs. This model clarifies the correlation between the adsorption energies and the intrinsic electronic and geometric properties of substrates and adsorbates, demonstrating that the electronic effects of active centers on adsorption are highly localized even down to the adsorption sites on SAAs, ABs and HEAs. Moreover, this predictive model screens out some potential alloy catalysts such as Cu@Ag, CuAg and AgAuCuPtPd for $CO_2RR$ with the onset potential of -0.47 V, -0.39 V and -0.59 V. Therefore, our ML scheme not only advances the comprehension of the adsorption mechanism on alloys, but also provides a convenient tool for the rapid design of alloy catalysts.

## 4. Methods

In this work, we implement the XGBoost regression in the code of the open-source xgboost package[14] combined with the Scikit-Learn API.[24] Hyperparameters are determined with the grid search method. Five-fold cross-validation is used to assess the performance of the ML model on an independent test set. We average the training model after 100 cycles for optimal performance and scalability on test sets. Performance metrics used in this work are the MAE and RMSE.

DFT calculations are conducted with the software package Vienna Ab-initio Simulation Package (VASP),[25] using Perdew-Burke-Ernzerhof exchange functional.[26] The kinetic energy cutoff for the wave function calculations is set at 500 eV. We use a four-layered (3×3) supercell for the (211) facet and a four-layered (4×4) supercell for (100) facet. A vacuum of at least 15 Å is adopted to separate the adjacent slabs. The Brillouin zones are sampled with 4×4×1 Monkhorst–Pack k-points for SAAs and ABs, as well as 2×2×1 for HEAs.

Moreover, the adsorption energy is defined as,

$$E_{ad} = E_{ads/sub} - E_{sub} - E_{ads}$$

where $E_{ads/sub}$ is the energy of the adsorbed system, $E_{ads}$ is the energy of an isolated molecule, and $E_{sub}$ is the energy of the clean metal substrate. For CO and OH, Cu(100) and Pt(111) is chosen as reference:

$$\Delta E_{ad} = \Delta E_{CO} - \Delta E_{CO/Cu(100)} = (E_{CO/sub} - E_{sub}) - (E_{CO/Cu(100)} - E_{Cu(100)})$$

$$\Delta E_{ad} = \Delta E_{OH} - \Delta E_{OH/Pt(111)} = (E_{OH/sub} - E_{sub}) - (E_{OH/Pt(111)} - E_{Pt(111)})$$

The free energy of adsorbates (*G*) is determined by incorporating corrections into the electronic energy calculated by DFT methods and is calculated as:

$$G = E_{DFT} + ZPE - TS$$

where $E_{DFT}$, ZPE and $S$ are the electronic energy, zero-point energy and entropy, respectively. Zero-point energy and entropy correction are calculated from vibrational frequencies analysis with harmonic oscillator approximation with the ideal gas approximation at 298.15 K.[27] The entropies of gas-phase molecules are obtained from the reference, which are consistent with the standard thermodynamic tables[28].

**Supporting Information**

Supporting Information is available from the Wiley Online Library or from the author.

**Acknowledgements**

We are thankful for the National Natural Science Foundation of China (Numbers 21673095, 11974128, and 51631004), the Opening Project of State Key Laboratory of High Performance Ceramics and Superfine Microstructure (SKL201910SIC), the Program of Innovative Research Team (in Science and Technology) in the University of Jilin Province, the Program for JLU (Jilin University) Science and Technology Innovative Research Team (Number 2017TD-09), and the Fundamental Research Funds for the Central Universities, and the computing resources of the High Performance Computing Center of Jilin University, China.

Received: ((will be filled in by the editorial staff))
Revised: ((will be filled in by the editorial staff))
Published online: ((will be filled in by the editorial staff))